# Title: Measuring Human Leadership Skills with Artificially Intelligent Agents

**Authors:** Ben Weidmann[1]*, Yixian Xu[1], David J. Deming[1]

**Abstract:** We show that the ability to lead groups of humans is predicted by leadership skill with artificially intelligent (AI) agents. In a large pre-registered lab experiment, human leaders worked with AI agents to solve problems. Their performance on this "AI leadership test" was strongly correlated with their causal impact on human teams, which we estimate by repeatedly randomly assigning leaders to groups of human followers and measuring team performance. Successful leaders of both humans and AI agents ask more questions and engage in more conversational turn-taking; they score higher on measures of social intelligence, fluid intelligence, and decision-making skill, but do not differ in gender, age, ethnicity or education. Our findings indicate that AI agents can be effective proxies for human participants in social experiments, which greatly simplifies the measurement of leadership and teamwork skills.

**Affiliations:**

[1]Harvard Kennedy School, Harvard University; Cambridge, 02138, USA

*Corresponding author. Email: benweidmann@hks.harvard.edu

**Main**

Good leadership is an important determinant of firm productivity and national prosperity (*1–4*). While many studies show the value of good leadership, we know relatively little about how to measure individual differences in leadership skills. Recent methodological developments have pointed towards the possibility of performance-based measures of leadership and teamwork skills (*5*, *6*). These measures identify the causal contribution individuals make to teams by observing people working in multiple, randomly assigned groups. This method reveals large differences in the contributions people make to team success, unrelated to characteristics such as gender, ethnicity, age and personality. While conceptually promising, the approach is costly and logistically challenging.

This paper asks whether agents created with large language models (LLMs) can be used to effectively simulate human interaction in team settings. In a large pre-registered lab study[1] we demonstrate that LLM agents can be used to measure individual differences in leadership skills. In the experiment, human leaders solve a series of collaborative problems by directing teams of AI agents. We compare performance on this 'AI leadership test' to a ground truth, in which we estimate each leader's causal contribution to human teams by repeatedly randomly assigning leaders to multiple teams of human followers. Our work builds on recent studies demonstrating that LLM agents can effectively approximate typical human responses (*7–12*).

Figure 1 provides an outline of our experimental design. The figure shows how the 'ground truth' assessment is conducted: leaders are randomly assigned to 6 different teams of human followers. Each time a leader is assigned to a group we make a prediction about group performance based on the individual skills of the leader and the followers (these individual skills are assessed at the very beginning of the experiment and include a measure of task-specific skill, along with fluid intelligence). Using these predictions we identify leaders whose human groups consistently outperform expectations. We then compare this 'ground truth' measure to performance on the AI leadership test. The AI test is analogous to the ground truth, but with AI agents replacing human followers. We balance the order of the assessments across leaders to ensure that our results aren't driven by practice effects.

The group task we use is a modified version of the 'Hidden Profile' problem, a widely-used social science paradigm for studying group decision-making (*13–16*); see Figure 2 for details. In hidden profile puzzles, each group member receives multiple pieces of information – some of which are known only to them. Groups need to surface information through conversation and synthesize their collective knowledge into a decision. This is a crucial component of successful teamwork (*16*) and requires several uniquely human capabilities such as flexible linguistic communication and perspective taking (*17*).

We modify the standard Hidden Profile problem in two ways. First, we create a distinct role for leaders, who are responsible for gathering information from the group and making a final decision. Second, we make the problem more difficult and realistic by adding probabilistic answers. For example, the correct solution may be that the answer is definitely not option A or B, but could be either option C or D, each with a 50% probability.

---

[1] https://aspredicted.org/g2pp-t8qv.pdf

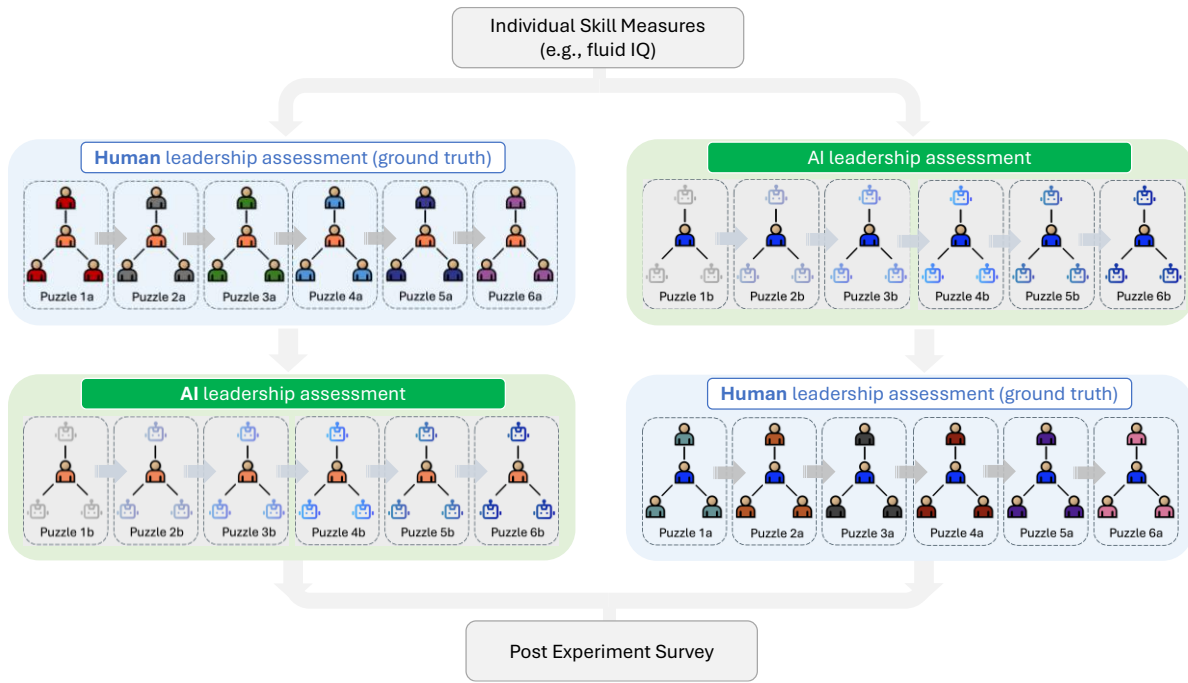

**Fig. 1. Overview of the experiment.** Each participant begins by completing a series of individual tasks, including a test of fluid intelligence (*18*); emotional perceptiveness (*19*); an individual analogue of the hidden-profile problem (see Methods section). Half the leaders complete 6 group puzzles with human teammates followed by 6 group puzzles with AI teammates. The other half of the leader sample complete 6 group puzzles with AI teammates, followed by 6 group puzzles with human teammates. All leaders then complete a post-experiment survey. Details of data collection logistics are provided in Supplementary Materials.

Our implementation involves a leader and three followers. The team is presented with a puzzle, for example diagnosing the cause of a faulty machine. The team has sufficient information to solve the puzzle, but essential clues are dispersed among the group such that no individual can solve the problem alone (*13*, *20*). Success depends on the team's ability to pool their unique, unshared information and make informed decisions (*21*). In our implementation the leader plays a vital role in this process: they ask questions and gather information from their team, manage their team's time such that they can address all relevant options, and synthesize the team's knowledge into a final decision.

Figure 2 presents an overview of our implementation and example clues and dialogue. The leader knows some but not all of the clues and can only arrive at a full solution through conversation with followers. The final answers – covering options A through E – are expressed as percent chances.

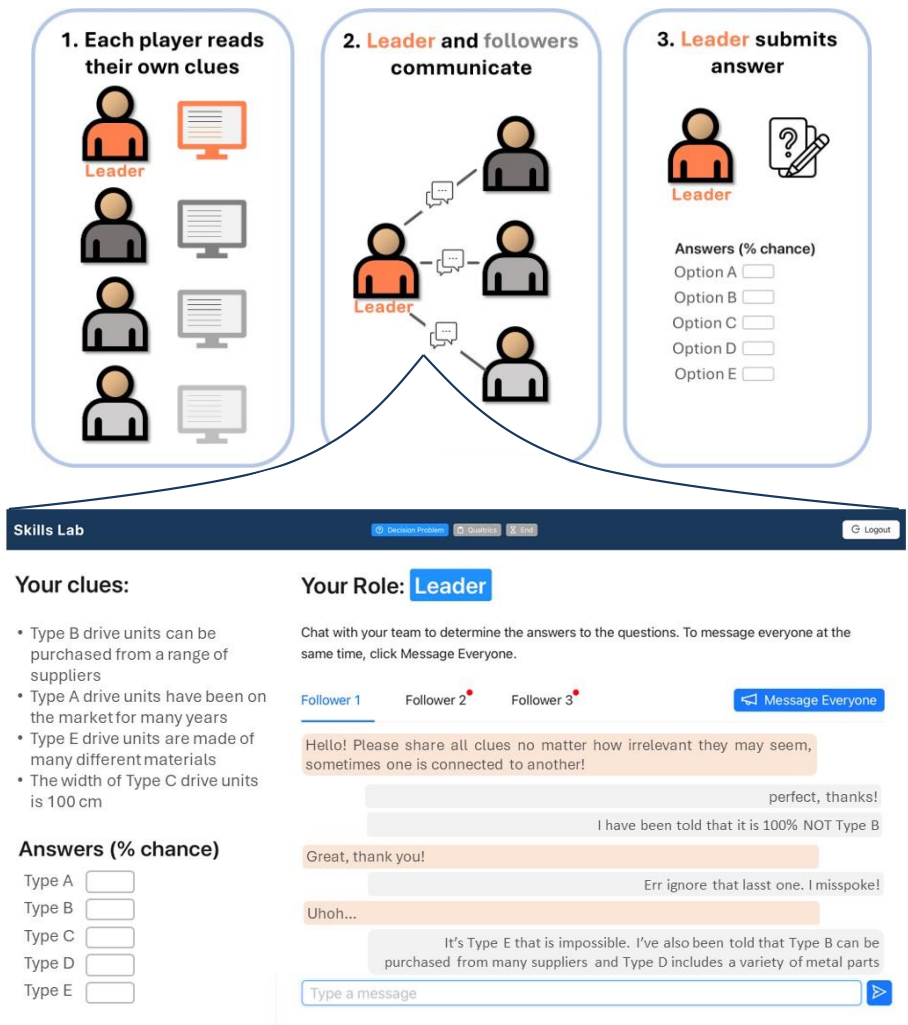

**Fig. 2. Overview of Group Task.** The figure provides an overview of the Hidden Profile task. The task proceeds in three main steps, as shown in the top row of the panel: i) participants are given information, some of which are shared among participants and some of which are private; ii) leaders then discuss their talk to their team in order to gather information and solve the puzzle; iii) leaders then indicate their posterior beliefs about a set of pre-defined answer options. The bottom pane of the figure illustrates the task interface. In the experiment, each puzzle contains two questions, but the figure only shows one for simplicity of exposition. Materials of group tasks are provided in Supplementary Materials.

The group task was created for the purpose of this study. We generated two parallel tests, one for human-only teams, and one for AI teams led by a human. Each test had the equivalent structure and difficulty across items, enabling fair comparisons between the AI test and the human test. By creating new puzzles, we also ensured that the AI agents would not have been exposed to the puzzle solutions in their training data.

Overall, our experimental setup allowed us to test three core, pre-registered predictions about the similarity of the AI Leadership Test with the ground truth test. First, that the magnitude of leaders' causal contributions will be the same in both tests. Second, that the same leader characteristics – in terms of demography and other skills – will predict success on both tests, to the same extent. Third, and most important, that individual leaders who succeed/(fail) in the AI test will also succeed/(fail) in the human-only test.

**Results**

This section starts by presenting results from our pre-registered hypotheses. We show that:
1. Variation in leader performance has a large and statistically significant impact on group performance, for both the AI test and the human ground truth.
2. The same skills and demographic factors predict success on both tests.
3. Individual scores on the AI test correlate very highly with ground-truth scores. This is true for raw scores and leadership measures that control for hard skills.

We then move to exploratory analysis and show that:
4. The same behavioral patterns predict success in both tests.
5. The AI Leadership Test reproduces substantive findings about leadership emergence and performance

*1. Variation in leader performance matters for group performance in both tests*

We find strong evidence that 'leader effects' – i.e. the causal impact a typical leader has on team performance – are large for both the AI leadership test and the human test (see Method section for formal definition of leader effects). More than half the variation in group performance on both tests can be explained *solely* by the identity of the leader.[2]

Figure 3 shows how similar the 'leader effects' are for the AI test and the ground truth.[3] The y-axis represents the causal impact on group performance (in standard deviation units) of replacing an average leader with a 'good' leader that is 1 SD above average. Replacing an average leader with a good one improves team performance by around 0.65SD. This improvement could be because some leaders have better 'soft skills' (e.g. teamwork and communication skills). Equally, it might be the case that leaders who make a strong contribution are primarily contributing 'hard skills' (e.g. being fast at typing or being good at logic puzzles).

The right panel of Figure 3 shows that leader effects are both large (around 0.55 SDs) even after conditioning on leaders' hard skills, which were assessed before the group experiment. Importantly, the AI test and the human ground truth have narrow, overlapping 95% confidence intervals. This suggests that the AI and human leadership tests are assessing similar 'soft skills', not merely picking up differences in task ability.

The difference between 'good' and 'bad' leaders is substantial. A good leader (1SD above average) correctly solves 53% of hidden profile problems, whereas a bad leader (1SD below average) solves only 10% of problems correctly.[4] The difference in performance between 'good' and 'bad' leaders is explored in Figure 4, which shows typical response profiles. Each question has five answer options and leaders need to assign their credence across these options. In the example, the clues rule out options B, C and D – and leave options A and E equally likely. The correct response profile is shown in the right column of the figure. The first column shows the answer profile of a 'good' leader and the second column illustrates the answer profile of a 'bad' leader (1SD below average). *The* typical response of 'bad' leaders is quite close to uninformed guessing, suggesting that they learn relatively little from interacting with their team.

---

[2] The $R^2$ of a model with fixed effects for leaders is 0.57 with the AI leadership test as an outcome, and 0.50 for the human test.
[3] The 'typical leader effects' are defined by parameter is $\hat{\sigma}_\alpha$, which is formally described in the Methods section.
[4] This averages across the AI and human measurements. Each Hidden Profile puzzle has two dimensions. 'Optimal' answers are defined at the dimension level.

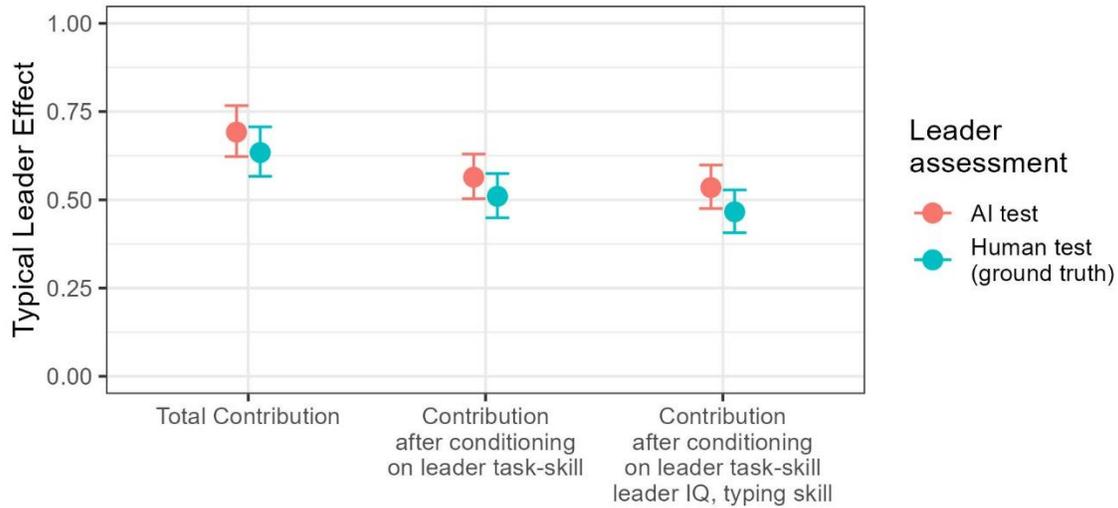

**Fig. 3. Magnitude of typical leader effects.** y-axis shows the typical leader effect ($\hat{\sigma}_\alpha$) defined in model (3) of the Method section. This represents the impact of a 1SD shift in leader quality on (standardized) group performance. Error-bars show 95% confidence intervals from profile likelihood inference. On the x-axis the 'Total Contribution' shows the magnitude of typical leader effects without conditioning on leader's task-specific skills or fluid IQ. The middle tick-mark represents estimates of the typical leader contribution after conditioning group scores on estimates of leader's individual task skills. The right tick-mark represents estimates of the typical leader contribution after conditioning group scores on leader's task skills, fluid IQ and typing speed. Note that leader individual task skills, fluid IQ and typing speed were assessed before the group assessments.

|  | GOOD Leader | BAD Leader | GUESS (flat prior) | CORRECT ANSWER |
|---|---|---|---|---|
| Option A | 50 | 28 | 20 | 50 |
| Option B | 17 | 18 | 20 | 0 |
| Option C | 0 | 18 | 20 | 0 |
| Option D | 0 | 18 | 20 | 0 |
| Option E | 33 | 18 | 20 | 50 |

**Fig. 4. Illustrative responses from 'good' and 'bad' leaders.** Each question has 5 answer options, and leaders need to submit credence that sum to 100. In this example options B, C, and D are ruled out by the clues given to participants, and options A and E are equally likely. 'Good leaders' have performance 1SD above average. 'Bad leaders' have performance 1SD below average. Items are constructed so that prior believes should be uninformative, so that leaders who learn nothing from the puzzles should submit the profile of an 'uniformed guess'.

## *2. The same skills/characteristics predict success on the AI test and the ground-truth*

If the AI Leadership Test does a good job of replicating the ground-truth test, we would expect that similar leader characteristics and skill profiles would predict success on both tests. This is what we find.

Figure 5 summarizes these results: the x-axis shows the correlation between various leader characteristics and success on the AI leadership test. The y-axis shows the correlation between the leader characteristics and success on the human test. The right panel shows the results for raw scores and the left panel shows scores conditioned on hard skills.

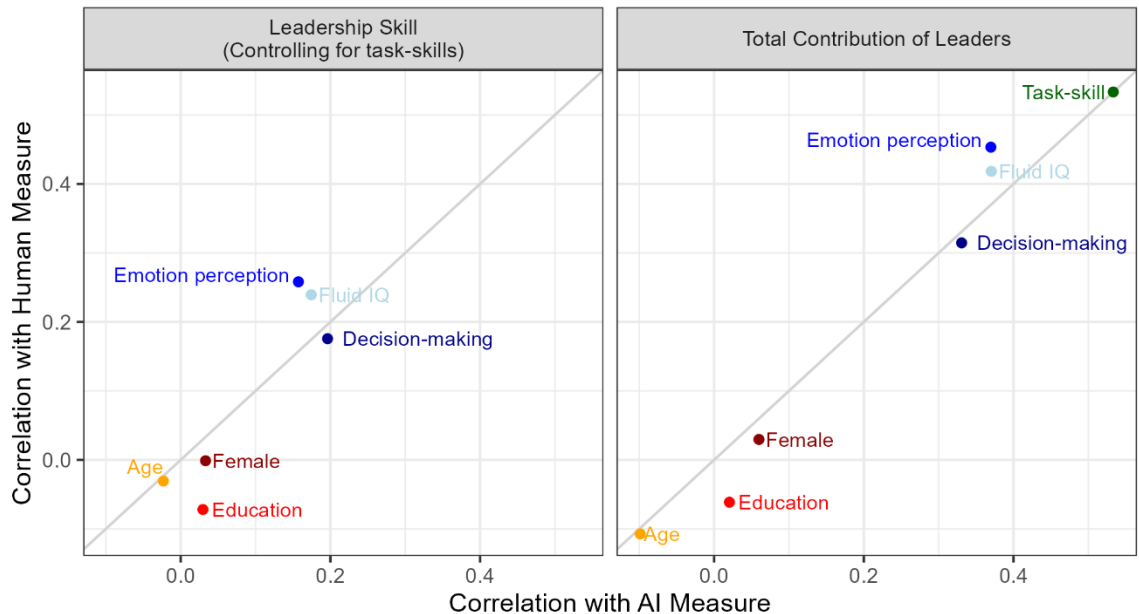

**Fig. 5. Predictors of success on the AI and Human measures.** The x-axis represents the correlation between a predictor variable X (e.g. 'Age') and performance on the AI test. The y-axis represents the correlation between covariates X and performance on the ground-truth measure. The left panel shows correlation coefficient for 'Leadership Skill', after group performance has been conditioned on task-specific skill i.e. cor(X, α); the right panel shows correlation coefficients for 'Total Contribution of Leaders', i.e. cor(X,Y). Each point represents two estimated correlation coefficients, with n=249 for each estimate.

The predictors of success are extremely similar across the AI and human-only tests. For both tests the strongest predictors of success are skill measures: fluid IQ, emotional perceptiveness, and a measure of economic decision-making.[5] Across both tests, demographic factors such as gender and ethnicity are not associated with leadership performance. The similarity of predictors holds regardless of whether leader contributions are conditioned to exclude the role of hard skills (left pane of Figure 5) or not (right panel).

One minor difference is the importance of emotional perceptiveness, which is more strongly associated with the human test. The correlation between emotional perceptiveness and success on the human leadership test = 0.45 (p<0.001), whereas for the AI test the correlation is 0.37 (p<0.001). This difference in correlation coefficients is not itself significant. Importantly, after conditioning on fluid IQ, the AI measure of leader contribution is still predicted by emotional perceptiveness (correlation = 0.24, p<0.001). Again, this suggests that both the AI test and the human test are measuring leadership skills beyond just individual differences in 'hard skills'.[6]

---

[5] The strongest predictor of each leader's total causal contribution to group success is their performance on the individual analogue of the group task, i.e. an individual hidden profile test, labelled 'task skill'. This is as expected: the individual hidden profile test was specifically designed to assess the hard skills needed to succeed in the group hidden profile test.

### 3. Individual scores on the AI test correlate very highly with ground-truth scores

We examine each leader's average raw score on the AI leadership test ($Y_i^{AI}$) and their average score across groups on the human test ($Y_i^{Human}$). At the level of individual leaders, the disattenuated correlation is $\hat{\rho} = 0.81$ (n=249), with a 95% confidence interval of [0.72,88].[7] In other words, individual scores on the AI Leadership test are very highly predictive of the total causal contribution that leaders make to groups.

As noted above, individual leaders can improve group performance through 'hard skills' and 'soft skills'. A higher bar for the AI Leadership Test is to see whether it predicts causal contributions of groups *after conditioning on hard skills.* Let $\alpha_i$ denote the average causal contribution that leader $i$ makes after controlling for hard skills (see Methods for details). We find that the disattenuated correlation between $\alpha_i^{AI}$ and $\alpha_i^{Human}$ is 0.69 (n=249) with a 95% confidence interval of [0.57,0.81].[8] This suggests that, at the level of individual leaders, the AI Leadership test does a good job of capturing leadership-specific soft skills that matter for success in human-only teams.

### 4. The same behavioral patterns are associated with success in AI and Human tests

Next, we turn to exploratory analyses. We start by analyzing the communicative strategies that are associated with success on the AI and Human leadership tests. We explore five process metrics, measured separately for each leader on each test. We find that the volume of communication – measured by the number of words – is not associated with good leadership. On the contrary, good leaders ask more questions and their teams engage in more conversational turn-taking. Good leaders also tend to use plural pronouns, referring to 'we' and 'us'. This mirrors previous findings on successful leader communication (*22, 23*).

Table 1 summarizes these results and shows the marginal association at the leader level (n=248) between various measures of communication and leadership scores on the AI and human tests. The patterns of association are very similar across the AI and human tests. One notable difference is in the impact of positive affect which is robustly associated with team success on the human test, but not so with the AI measure. We return to this result in the discussion.

### 5. The AI Leadership Test reproduces substantive findings about leadership

Thus far, we've shown that individual scores on the AI Leadership Test are remarkably similar to the ground-truth scores. We now explore whether the AI test may also benefit social science by testing substantive hypotheses about leadership and teamwork dynamics. Here, we extend a fast-growing literature examining the use of 'silicon samples', i.e. using samples of LLMs to complement human studies (*24, 25*).

---

[6] Similarly, the association between emotional perceptiveness and leader contributions is robust to conditioning on task-specific skill for both the AI test (correlation = 0.16, p<0.02) and the human test (correlation = 0.26, p<0.001). Note that the difference between these correlations is not significant.
[7] Raw correlation = 0.67.
[8] Raw correlation = 0.52.

We first examine people's willingness to take leadership roles. Lack of appetite for leadership positions has been cited as a concern in practice (*26*) especially in terms of gender equity (*27*). We test whether overconfident people are more willing to be leaders. We measure overconfidence by asking leaders to rate their performance relative to their peers and comparing this to their actual performance.[9] We measure overconfidence separately for the human measure *and* the AI measure.[10] We find that participants' willingness to lead a team of AI agents is positively associated with overconfidence ($\hat{\rho} = 0.29$, p<0.005). This reproduces the result from the human measures, which finds a correlation between overconfidence and willingness to lead human teams of $\hat{\rho} = 0.27$ (p<0.005).

Second, we study the association between the accuracy of leaders' self-evaluations and the causal contribution that leaders make to their teams. We are unaware of any studies that examine this association, largely due to the difficulty of estimating leader causal contributions. As such, we think it very unlikely that our findings would be present in AI training data. We find that leaders who are more accurate in evaluating their performance *contribute more to their teams*. This association emerged independently for the AI measure ($\hat{\rho}=0.16$, p=0.08) and the human assessment ($\hat{\rho}=0.21$, p=0.02).

**Discussion and conclusion**

We view our implementation as a proof-of-concept for a practical, performance-based measure of leadership skill. However, our work has several limitations. While the AI and human tests show remarkable similarities, they are not identical: the cognition and behavior of our AI agents differ from that of humans (*29*). We note that emotion appears to play a diminished role in the AI Leadership Test. This is evident in two ways: there is an association between leaders' use of positive affect and group performance in the human test, but not when the followers are LLM agents. Similarly, the association between emotional perceptiveness and leader performance is slightly weaker in the AI test, compared with the human analogue. Other behavioral differences were reported by some leaders in their qualitative feedback, including the observation that strong human followers were 'better to work with' than AI followers in terms of filtering information and occasionally providing meta-level advice about successful puzzle-solving strategies.

A second broad class of limitations is that our implementation does not attempt to replicate the variety of human behavior in AI agents. This represents an important direction for future research. The current state of large language models enables considerable customization of voice and behavior. In future work we hope to use data from human experiments to help silicon samples better reflect the rich diversity of human behavior (*8*).

Finally, a crucial next step will be to externally validate our approach by linking assessment outcomes to leadership success in real-world settings. This validation will be important for establishing the practical utility of our method.

With these limitations in mind, we believe that our approach to leadership measurement makes contributions on three fronts. First, our method offers a path to improving how leaders and managers are selected in the labor market. Our results suggest that the causal impact leaders have on groups is not strongly predicted by demographic factors, mirroring findings from the field (*8*). This suggests that current selection processes are overlooking many high-potential leaders. A low-cost, standardized assessment could therefore improve both organization performance and social equity.

---

[9] In the terminology of Moore and Healy (*28*) this is a measure of 'overplacement'.
[10] Willingness to lead is measured by asking participants to rate, on a scale of 1-10, their appetite for being a leader in a future experiment. Again, we measure this separately for the AI leadership measure, and the human test.

Second, practical and scalable leadership assessments may increase the total supply of leadership skills in the workforce by enabling leadership educators to better evaluate their programs. Despite compelling evidence about the importance of leadership skills, most leaders receive no formal training (*33*) – and the training that exists is rarely subject to robust, causal evaluation (*34*). We believe that the absence of robust performance-based tests is a barrier to evaluation, which may be lowered using AI-based tests.

Last, our approach can potentially enable more efficient testing of hypotheses about leadership and teamwork skills. The contrast in resources required for the human versus AI versions of our test is telling: while the human version cost $114 per participant and required oversight from two researchers, the AI version cost $23 and ran autonomously. More importantly, the AI version did *not* require us to coordinate a large group of people to be available at the same time, as is the case for human-only group assessments using repeated randomization. This substantial reduction in costs and logistical complexity could increase the number of scholars who can pursue leadership and teamwork research.

In conclusion, our study demonstrates that an AI-based test of leadership skills corresponds remarkably closely with a ground-truth measure where humans lead teams of humans. Both tests are strongly predicted by the same broad skills and communication strategies. This alignment suggests that AI-based assessments could offer scalable, standardized, performance-based measures of 'soft skills'. Evidence suggests that these skills are increasingly valued by the labor market (*35*, *36*) and this paper highlights a possible path to improve the measurement and supply of these crucial skills.

**Method**

*1. Participant recruitment and flow*

The core of our experiment is a collaborative problem-solving task based on the Hidden Profile paradigm (*16*). The task, described in detail in the next section, measures each leader's ability to gather information and make decisions. Every leader solves a series of Hidden Profile problems in two conditions: with human followers ('human test') and with AI followers ('AI test'). In each condition, leaders work with 6 groups. To account for potential practice effects, the order of the two conditions is counterbalanced across participants: roughly half the leaders complete the human test first (n=121) while the other half begin with the AI test (n=128). Figure 6 illustrates the flow of participants through our experiment.

All participants begin the experiment by completing a series of individual tests (described in '3. Individual assessments'). Table 2 shows the balance across leaders who completed the tests in different orders. Participants who completed the human test first appear to have a small advantage in terms of task skill and emotional perceptiveness, but overall the two groups are well balanced. LLM followers were generated using GPT4o.[11]

---
[11] The prompt for the LLM is available in the Methods section in Supplementary Materials, and is analogous to the instructions human participants received before working on the group task.

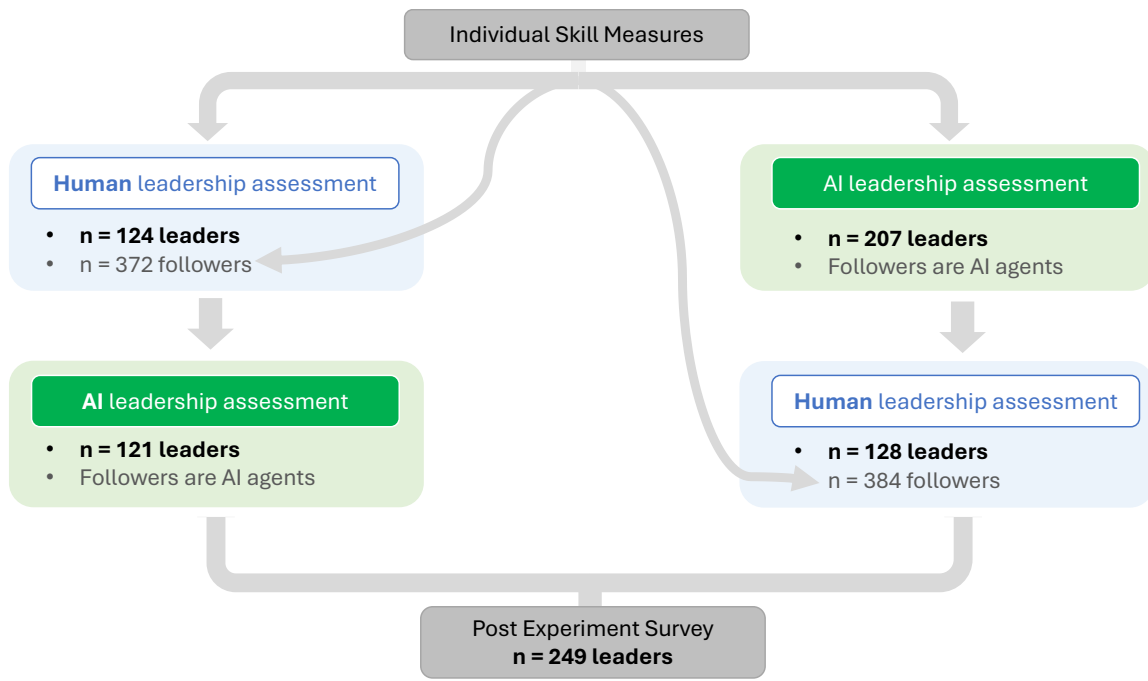

**Fig. 6. Flow of participants through the experiment.** The figure shows the flow of participants across conditions, including the number of leaders and followers per group session. The longitudinal experiment has four parts. Individual skill measures and the first session of group task (Human or AI leadership assessment) take place on the same day. The second group session, featuring the complementary leadership assessment happens 1–2 days later. This is followed immediately by the post-experiment survey. Details of data collection logistics are provided in Supplementary Materials.

## *2. Group task: Hidden Profile*

Each group in the experiment worked collaboratively on a Hidden Profile problem (task materials available in Supplementary Materials). Each problem is self-contained and has an objectively correct solution. The core feature of Hidden Profiles is that the information needed to arrive at the solution is distributed among team members, such that no individual can solve the problem alone (*13*, *20*). Success depends on the team's ability to effectively pool their unique, unshared information and use the information to make informed decisions (*21*).

Our task is similar to other online experiments using Hidden Profiles (*37*). Each group is presented with a scenario, such as troubleshooting a broken machine. In each scenario, groups need to answer two questions. For example, in the 'strange fish' scenario, participants need to classify a type of rare fish, and diagnose which virus it's suffering from. Each question has five pre-determined options (e.g. the fish might be 'Blackfish'; 'Bluefish'; 'Redfish'; 'Yellowfish' or 'Greenfish'). The puzzles required no outside knowledge and answer options were chosen such that participants would be indifferent to them before reading the clues. Each team member received 8 clues per puzzle, including 4 clues for each dimension. The information distribution consisted of half public and half private clues for every question. Clues are classified as either 'disqualifying' or 'distractor' clues. 'Disqualifying' clues rule out specific options (e.g., 'Report #7 indicates the fish does not have yellow spotted scales so it's not the Yellowfish') while 'distractor' clues present information that appears to be factual but is irrelevant to the problem at hand (e.g., 'Report #5 indicates that Greenfish migrate to access seasonal food resources').

Each puzzle was newly created for the purpose of this study. This allowed us to generate two parallel forms of Hidden Profile problems, each with equivalent structure and difficulty across items. These parallel forms enabled fair comparisons between the AI test and the human test. By creating new puzzles, we also ensured that the AI agents would not have been exposed to the puzzle solutions in their training data.

Creating new Hidden Profile problems also allowed us to modify traditional implementations in two important ways. First, we introduce hierarchy in order to study leadership skills. Each group has one leader and three followers (either three humans, or three LLM agents). The leader has several distinct responsibilities: gathering information from their team; managing the team's time such that they can address all relevant options; and synthesizing the team's knowledge into a final decision. Group communication is arranged in a star network, such that the leader can communicate with anyone in the group, but followers can only talk to the leader.

The second modification is to require participants to answer probabilistically, rather than simply trying to identify a single correct candidate. In some cases, even with perfect communication groups can only narrow potential answers down to 3 or 4 of the pre-specified options. In these cases, leaders need to divide their credence across different options. Participants are familiarized with this setup in Part 1 during the individual Hidden Profile problems. Moving from 'single correct answer option' to 'probabilistic solutions' reduces issues with ceiling effects (*37*). We score each participant's responses by calculating how closely their assigned probabilities for each option match the correct answers. The smaller the discrepancy, the higher the score. Compared with a binary score of 'correct'/'incorrect', this continuous scoring method provides more information about participants' ability on the task and reduces measurement error.

### *3. Individual assessments*

#### *3.1 Measuring 'Hard Skills'*

The group hidden profile task requires leaders and followers to have a set of concrete, hard skills. For example, as groups communicate via text-based chat and the task has a time limit, leaders who are skilled a typing likely have an advantage. In identifying the contribution that leaders make to their team, our goal is to decompose the contribution of task-specific hard skills (such as typing skill) from more general leadership skills, such as asking effective questions, generating positive affect among teammates, and so on. This decomposition requires us to measure the 'hard skills' leaders may have that contribute to team success. We focus on three measures.

*Typing Task:* we measure participants' typing skills by giving them 1 minute to type as much of a news article as they can. Performance was scored by the number of correctly typed words.

*Individual Hidden Profile Task:* before the group task, we assess participants' ability to solve an individual-analogue of 'hidden profile' puzzles. Participants work alone and are given all the information they need to provide a correct answer. The task assesses how well participants synthesize a corpus of clues into probabilistic inferences. The task structure matches the group task: participants receive a set of clues to solve a puzzle, with 5 pre-determined answers. Some clues definitively rule out particular options, while others are distractors and provide no useful information. Each participant completed three individual hidden profile puzzles. The task was purpose-built for this experiment and screenshots of the interface are available in fig. S1 of Supplementary Materials.

*Fluid IQ:* we measure fluid IQ using the Culture Fair Intelligence Test (CFIT III), a standard measure for problem-solving ability. An example item of the test is provided in panel A of Figure 7.

*3.2 Other individual skill measures*

*Economic Decision-Making Skill*: we measured economic decision-making skill, defined by Caplin et al. (*36*) as the ability to make good resource allocation decisions, using a short version of the Assignment Game (*36*). In this task, participants act as managers, assigning workers to tasks. To perform well, participants must understand comparative advantage, and avoid decision-making biases such as anchoring in a complex numerical environment. A screenshot from the game is presented in panel C of Figure 7.

*Emotional Perceptiveness:* social intelligence was assessed using the PAGE (Perceiving AI Generated Emotions) test, a modern measure of emotion perception (*19*). Participants were presented with 35 images of diverse faces displaying 20 emotions and were asked to identify the correct emotion for each face, from six options. The PAGE test has better psychometric properties and demonstrated higher predictive validity in teamwork settings compared to the widely-used Reading the Mind in the Eyes Test (RMET). An example is presented in Figure 7, Panel B.

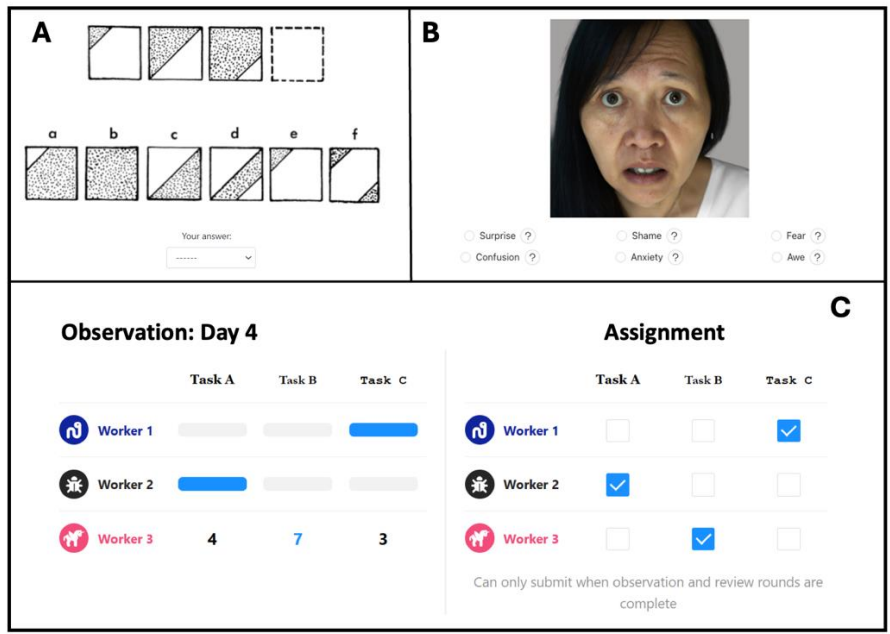

**Fig. 7. Individual skill assessments.** (**A**) Example item of the Culture Fair Intelligence Test (CFIT III) (*18*). (**B**) Example item of the emotion perception measure (*19*). (**C**) Example item of the Assignment Game, which measures economic decision-making skill (*36*). Materials of all individual tasks are available in Supplementary Materials.

*3.3 Post-Experimental Survey*

Finally, to understand leaders' experiences with human and AI followers, we conducted a post-experimental survey, which covered self-perceived leadership effectiveness, willingness to lead, and leadership strategies across two conditions.

## *4. Identification of leadership skills*

*4.1 'Ground truth'*

To identify the total causal contribution that leaders make in the human-only assessment, we exploit the random assignment of leaders to groups. To focus specifically on the contribution that leaders make *independent of their hard skills* we condition group performance on measures of leaders hard skills. Our approach follows previous research and was pre-specified.[12]

Let leaders be indexed by $i = 1, ..., n$. In the human ground-truth assessment, leaders are randomly assigned to groups of three people, with groups indexed by $g$. Let the performance of group $g$ be given by $G_g$ and let $I_{ig}$ be a binary indicator equal to 1 if leader $i$ is in group $g$. Because leaders are randomly assigned to teams the total average causal effect that leader $i$ on groups is simply the average of $G_g$ for groups led by leader $i$. [13]

Next, we attempt to isolate the causal impact leaders have on groups *independent* of their hard skills. We denote this impact as $\alpha_i$. To estimate the $a$ parameters we condition group scores on previously-assessed measures leader's hard skills, denoted by $X_i$

$$G_g = \gamma \sum_g I_{ig} X_i + \epsilon_{ig} \quad (1)$$

$$a_i = \frac{1}{\sum_g I_{ig}} \sum_g \hat{\epsilon}_{ig} \quad (2)$$

In our framework, $\hat{a}_i$ is an estimate of leader $i's$ average causal contribution, conditional on how skilled leader $i$ is in terms of task-relevant hard skills.

To estimate the typical magnitude of the leader contributions, we estimate a parameter labelled $\sigma_a$, which represents the standard deviation of $a_i$. In order to do this we fit a multilevel model[14]:

$$\hat{\epsilon}_{gi} = a_i + e_{gi} \quad (3)$$
$$a_i \sim N(0, \sigma_a^2)$$
$$e_{gi} \sim N(0, \sigma^2)$$

Confidence intervals for key parameters are estimated using Profile Likelihood.

*4.2 Estimating leadership skills in the AI assessment*

Following our pre-registration we replicate these analyses for our AI assessment. Once again, we can calculate the average score for leader $i$ directly as the average of $G_g$ for the groups led by $i$. We then estimate equations (1), (2) and (3) using data from the AI assessment to estimate $a_i^{AI}$ and $\sigma_a^{AI}$.

---

[12] https://aspredicted.org/g2pp-t8qv.pdf

[13] Formally, this is $\frac{1}{\sum_g I_{ig}} \sum_g G_g I_{ig}$

[14] As noted in Weidmann and Deming (*5*), a multilevel model is needed here as simply estimating the standard deviation of $\hat{a}$ estimates yields $\hat{\sigma}_{\hat{a}}$ which is biased upwards by measurement error.

|  | Human (1) | AI (2) | Human (3) | AI (4) | Human (5) | AI (6) | Human (7) | AI (8) | Human (9) | AI (10) | Human (11) | AI (12) |
|---|---|---|---|---|---|---|---|---|---|---|---|---|
| #Words | 0.094 (0.063) | 0.028 (0.065) |  |  |  |  |  |  |  |  | -0.052 (0.077) | -0.112 (0.075) |
| #Questions |  |  | 0.130** (0.063) | 0.150** (0.064) |  |  |  |  |  |  | 0.100* (0.060) | 0.104 (0.065) |
| #Turns |  |  |  |  | 0.202*** (0.062) | 0.175*** (0.064) |  |  |  |  | 0.216*** (0.073) | 0.187*** (0.069) |
| Plural Pronouns |  |  |  |  |  |  | 0.172*** (0.063) | 0.156** (0.064) |  |  | 0.169*** (0.063) | 0.142** (0.068) |
| Positive Affect |  |  |  |  |  |  |  |  | 0.245*** (0.062) | 0.115* (0.064) | 0.245*** (0.060) | 0.123* (0.064) |
| Observations | 248 | 244 | 248 | 244 | 248 | 244 | 248 | 244 | 248 | 244 | 248 | 244 |
| $R^2$ | 0.009 | 0.001 | 0.017 | 0.022 | 0.041 | 0.030 | 0.030 | 0.024 | 0.060 | 0.013 | 0.141 | 0.081 |

**Table 1. Communication predictors of leader performance.** The table presents 12 regressions, all at the level of individual leaders. For the columns labelled 'AI' the dependent variable is the causal contribution leader i in the AI leadership test, conditioning on leader's task-skills (our pre-registered measure). For the columns labelled 'human' the dependent variable is the causal contribution for leader i in human test (also conditioned on task-skills, as per our pre-registration). Both outcome variables have been standardized to have Mean=0 and SD=1. We examine five predictor variables, all of which are standardized within each test to have mean=0 and SD=1. See Methods in Supplementary Materials for measurement details. **p<0.05, ***p<0.01.

|  | Human-test first | AI test first | p-val |
|---|---|---|---|
| Female (%) | 57% | 46% | 0.10 |
| Age (mean, years) | 38.4 | 36.8 | 0.14 |
| Attended university (%) | 74% | 74% | 0.99 |
| Ethnicity white (%) | 60% | 59% | 0.82 |
| Ethnicity black (%) | 21% | 25% | 0.44 |
| Ethnicity asian (%) | 7% | 5% | 0.50 |
| Ethnicity mixed, other or prefer not to say (%) | 12% | 11% | 0.51 |
| Task skill (mean score on individual hidden profile analogue) | 0.71 | 0.65 | 0.01 |
| Fluid IQ (mean score on Ravens) | 5.0 | 4.8 | 0.69 |
| Economic decision-making (mean score on Assignment Game) | 20.8 | 19.2 | 0.16 |
| Emotional perceptiveness (mean score on PAGE) | 24.5 | 23.0 | 0.01 |
| n | 121 | 128 |  |

**Table 2. Sample statistics.** Sample statistics for leaders. Column 2 shows characteristics of leaders who completed the human leadership test first followed by the AI test. The leaders in column 3 had the order of the tests reversed.

**Acknowledgments:** the authors thank Cooper Reed and Humberto Evans for their excellent technical contribution to the experiment.

**Funding:** the authors gratefully acknowledge the financial support of the Walmart Foundation.

**Author contributions:**
Conceptualization: BW, YX, DD
Methodology: BW, YX
Investigation: YX
Project administration: YX, BW
Analysis: BW, YX
Writing: BW, DD, YX
Funding acquisition: DD, BW

**Competing interests:** The authors declare that they have no competing interests.

**Data and materials availability:** All data, code, and materials are available in a publicly accessible OSF repository (DOI 10.17605/OSF.IO/QDY9V). The GPT prompts used to condition the AI agents, experiment instructions, and screenshots of all task interface are provided in the supplementary materials.